\documentclass{epsconf}
\usepackage{graphicx}
\usepackage{wrapfig}
\usepackage{amsmath}

\usepackage{hyperref} 
\hypersetup{colorlinks = true,linkcolor = blue, breaklinks = true}

\title{New solution to Airy's equation for modeling beams near turning points}
\author{\underline{N.~A.~Lopez}$^1$}
\institute{$^1$ Rudolf Peierls Centre for Theoretical Physics, University of Oxford,
Oxford OX1 3PU, UK}

\begin{document}
\maketitle

\section{Introduction}

The accurate computation of wavefields near reflection points is both intrinsically interesting from a basic science point of view and practically important for developing thermonuclear fusion as a future clean energy source. For example, reflection physics may contribute to glint losses in hohlraums~\cite{Lemos22} or trigger nonlinear instabilities~\cite{Turnbull15}. Many current approaches assume the wavefield to have an Airy profile transverse to the reflecting surface; however, this solution is only strictly true for incident plane waves. Since wavefields used in applications are not plane waves, a detailed investigation of the reflection profiles for general wavefields is needed.

Here I summarize the results presented in more detail elsewhere~\cite{Lopez23}. In particular, I show that when asymptotic matching to a prescribed incident wavefield is performed, the solution changes from being the Airy function to being more generally the hyperbolic umbilic function~\cite{Olver12, Poston96}. This solution accounts for (de-)focusing of the incident wavefield due to gradient-index lensing in addition to the familiar Airy physics, which importantly do not superimpose. I shall restrict attention to normal incidence with no spatial damping; Ref.~\cite{Lopez23} presents the more general case.

\subsection{General theory}

Consider a 2-D electromagnetic wave propagating in a plasma density profile that depends linearly on $x$. For normal incidence, the wavefield satisfies the Helmholtz equation
\begin{equation}
	\left(
		\partial_x^2 
		+ \partial_y^2
		+
		\frac{L - x}{\delta^3}
	\right)\psi(x, y) = 0
	, \quad
	\delta 
	\doteq \sqrt[3]{\frac{L \lambda^2}{4 \pi^2}}
	.
	\label{eq:airyEQ}
\end{equation}

\noindent Note that $\delta$ is the Airy skin depth; all other symbols have their usual meaning. By performing a Fourier transform in $y$, one can show that the solution takes the form
\begin{equation}
	\widetilde{\psi}(x, k_y)
	=
	\frac{
		\text{Ai}\left(
			\delta^2 \,k_y^2
			+ \frac{x - L}{\delta}
		\right)
	}{
		\text{Ai}\left(
			\delta^2 \,k_y^2
			- \frac{L}{\delta}
		\right)
	}
	\,
	\widetilde{\psi}(0, k_y)
	.
	\label{eq:airyFT}
\end{equation}

However, $\widetilde{\psi}(0, k_y)$ depends on both the incoming component (generally known) and the reflected component (generally unknown). Therefore, let us perform an asymptotic matching to construct a solution only in terms of the incident wavefield. If one has
\begin{equation}
	\frac{L}{\delta} - \delta^2 \,k_y^2
	\gg 1
\end{equation}

\noindent for all $k_y$ in the incident spectrum of $\psi$, then the turning point (which is shifted for finite $k_y$~\cite{Kruer03}) and the incident plane are sufficiently separated to allow the asymptotic form for $\text{Ai}$ to be used. This asymptotic form contains separate incoming and reflected waves, so one can isolate just the incoming component to conclude
\begin{equation}
	\widetilde{\psi}(x, k_y )
	=
	\sqrt{- 4\pi i} \, \sqrt[4]{\frac{L}{\delta} - \delta^2 \,k_y^2} \, 
	\text{Ai}\left( \delta^2 \,k_y^2 + \frac{x - L}{\delta} \right) 
	\exp\left[
		i \frac{2}{3}
		\left( \frac{L}{\delta} - \delta^2 \,k_y^2 \right)^{3/2}
	\right]
	\widetilde{\psi}_\text{in}(0, k_y)
	.
	\label{eq:asymIN}
\end{equation}

\noindent The solution is then obtained by inverse Fourier transform.

To make the nature of the caustic more apparent, one should further approximate Eq.~\ref{eq:asymIN} as
\begin{align}
	\sqrt[4]{\frac{L}{\delta} - \delta^2 \,k_y^2 }
	\approx \sqrt[6]{ \frac{2 \pi L}{\lambda} }
	, \, \quad
	\left( \frac{L}{\delta} - \delta^2 \,k_y^2 \right)^{3/2}
	\approx
	\frac{2 \pi L}{\lambda}
	- \frac{3 \lambda L}{4\pi} k_y^2
\end{align}

\noindent Then, applying the inverse Fourier transform to Eq.~\ref{eq:asymIN} yields
\begin{equation}
	\psi(x, y)
	\approx 
	N \,
	\psi_\text{in}( 0,y')
	*
	\text{U}_\text{H}
	\left(
		\sqrt[3]{3} \frac{x - L}{\delta}
		,
		\frac{ 
			y
		}{\sqrt[6]{3} \, \delta}
		,
		- \frac{\lambda L}{2\pi \sqrt[3]{3} \, \delta^2 }
	\right)
	,
	\label{eq:solUH}
\end{equation}

\noindent where $*$ denotes the convolutional product $f*g = \int \text{y}' f(y) g(y - y')$. I have also introduced the normalization constant
\begin{equation}
	N \doteq 
	\frac{\sqrt[6]{3}}{2\pi \sqrt{i \pi} \delta} 
	\sqrt[6]{ \frac{2 \pi L}{\lambda} }
	\exp\left(i \frac{4 \pi L}{3 \lambda} \right)
	.
\end{equation}

\noindent The function
\begin{equation}
	\text{U}_\text{H}(t_1, t_2, t_3)
	\doteq
	\int \text{d} u \, \text{d} v \,
	\exp(
		i u^2 \, v
		+ i v^3
		+ i t_3 \, u^2
		+ i t_2 u
		+ i t_1 v 
		)
	\label{eq:hyperUMB}
\end{equation}

\noindent is the standard $D_4^+$ hyperbolic umbilic function from catastrophe theory~\cite{Olver12,Poston96}. It contains the Airy function as a special case, but more generally it describes the combined effect of gradient-index lensing and the reflection interference. Importantly, because it is a catastrophe function, $\text{U}_\text{H}$ is structurally stable and therefore expected to describe the correct qualitative behavior even if the problem parameters change slightly (such as for oblique propagation or deviating slightly from the purely linear density profile). It should also be emphasized that Eq.~\ref{eq:solUH} is still an exact solution of Eq.~\ref{eq:airyEQ}; the approximation sign only applies to the mapping $\psi_\text{in} \mapsto \psi$.

\section{Plane wave special case}

Consider when $\psi_\text{in}$ is given by a plane wave
\begin{equation}
	\psi_\text{in}(0, y) = E_0
	.
\end{equation}

\noindent Then Eq.~\ref{eq:solUH} reduces to read
\begin{equation}
	\psi(x, y)
	=
	\frac{2 \pi E_0}{ \sqrt{i \pi}}
	\sqrt[6]{ \frac{2 \pi L}{\lambda} } \,
	\text{Ai}\left(
		\frac{x - L}{\delta}
	\right)
	\exp\left(
		i \frac{4 \pi L}{3 \lambda}
	\right)
	.
\end{equation}

\noindent Hence, the new general theory given by Eq.~\ref{eq:solUH} succesfully reproduces the standard Airy solution when the incident field is a plane wave, as desired. Note that the constant prefactors ensure that the initial field has amplitude $E_0$.

\section{Gaussian-focused wave special case}

Next, consider when $\psi_\text{in}$ is given by a Gaussian-focused wave
\begin{equation}
	\psi_\text{in}(0,y) = 
	E_0
	\exp\left( 
		- i \frac{\pi y^2}{\lambda f}
	\right)
	,
\end{equation}

\noindent with $f$ being the focal length. (Note that a Gaussian-focused beam corresponds to a Gaussian-focused wave with complex focal length~\cite{Lopez23}.) Equation~\ref{eq:solUH} then yields
\begin{equation}
	\psi(x, y)
	=
	E_0 
	\frac{ \sqrt[6]{3} \sqrt{\lambda f }  }{2\pi i \sqrt{\pi} \delta}
	\sqrt[6]{ \frac{2 \pi L}{\lambda} } \,
	\exp\left(
		i \frac{4 \pi L}{3 \lambda} 
	\right)
	\text{U}_\text{H}
	\left(
		\sqrt[3]{3} \frac{x - L}{\delta},
		\frac{ 
			y
		}{\sqrt[6]{3} \, \delta},
		\lambda \frac{
			f
			- 2 L
		}{4 \pi \sqrt[3]{3} \, \delta^2}
	\right)
	.
	\label{eq:solGAUSS}
\end{equation}

\noindent In other words, a Gaussian-focused incident field yields the hyperbolic umbilic function as the total solution. It is through analyzing this correspondence that one can gain intuition for the hyperbolic umbilic function in Eq.~\ref{eq:solUH}.

\begin{wrapfigure}{r}{70mm}\centering
\vspace{-1cm} 
\includegraphics[width=70mm, trim = 4mm 10mm 5mm 2mm, clip]{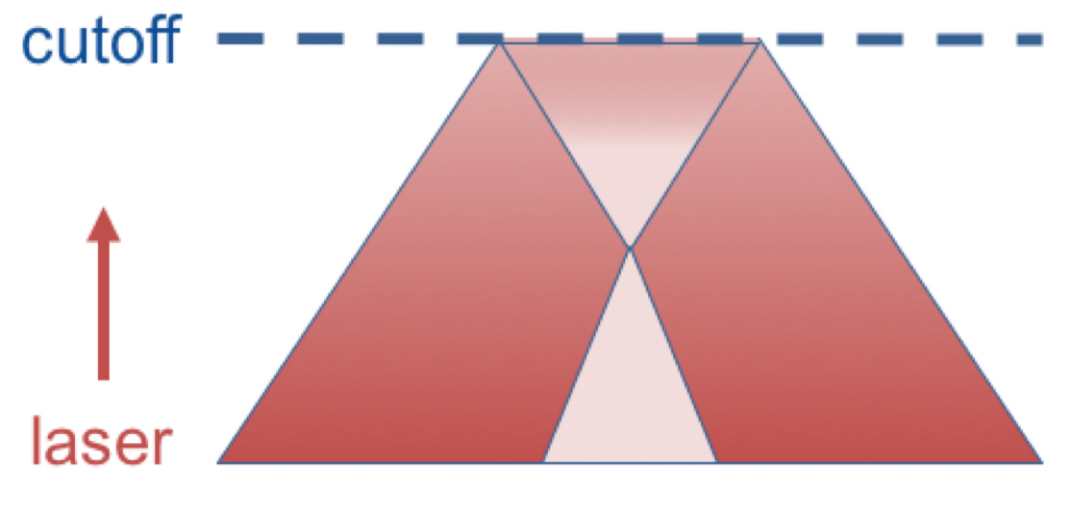}

\vspace{1mm}
\includegraphics[width=70mm, trim = 3mm 4mm 3mm 5mm, clip]{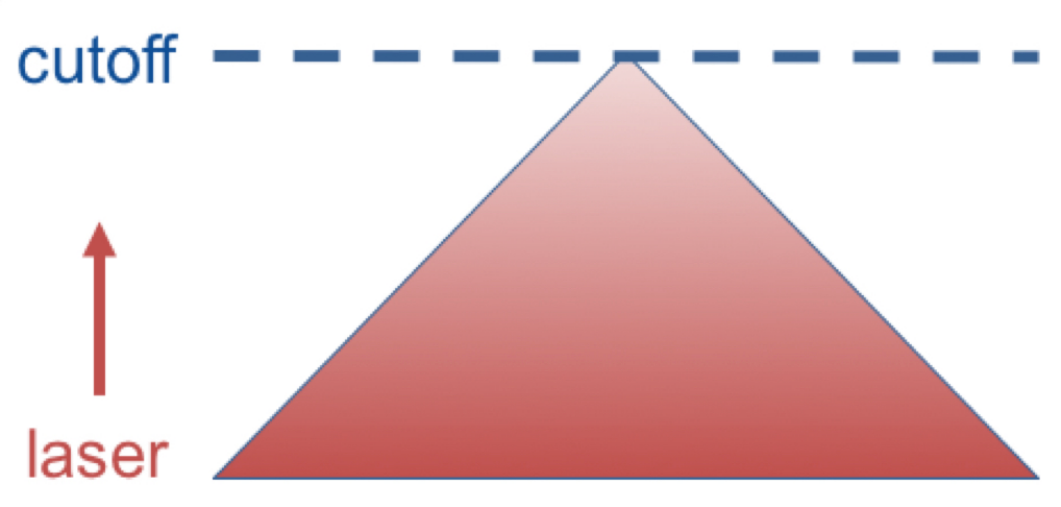}
\caption{\it \small Cartoon depicting the morphology of $\text{U}_\text{H}$. When $E$ is focused far from the turning point (top), the focusing and the reflection physics decouple to yield an Airy pattern near the turning point. If $E$ is critically focused (bottom), however, the reflection physics is more complicated. See Ref.~\cite{Lopez23} for more.}
\label{fig:cartoon}
\vspace{0.5cm} 
\end{wrapfigure}

For example, Eq.~\ref{eq:solGAUSS} predicts a critical point at the specific focal length $f = 2 L$, when the gradient-index lensing of the linear plasma density conspires to sharply focus the wavefield at the turning point. This corresponds to the most singular behavior of $\text{U}_\text{H}$ and can exceed the intensity peak of the Airy function by an arbitrarily large amount. Indeed, the ratio of the intensity peaks
\begin{equation}
	\frac{
		|\text{max} \text{U}_\text{H} |^2
	}{ 
		|\text{max} \text{Ai} |^2
	} 
	\sim \left( \frac{\lambda}{L} \right)^{-1/3}
\end{equation}

\noindent diverges in the geometrical-optics limit $\lambda/L \to 0$. This has consequences for setting intensity limiters for ray-tracing simulations near turning points. For example, a 351 nm wave incident on a plasma with 1 mm density scale length can potentially swell $14$ times higher than the Airy limiter often used~\cite{Myatt17}. That said, since a Gaussian-focused wave approaches a plane wave in the limit $f \to \infty$ it must be the case that $\text{U}_\text{H}$ reduces to $\text{Ai}$ when the focusing terms are negligible close to the turning point. The general morphology of Eq.~\ref{eq:solGAUSS} is given by the series of cartoons in Fig.~\ref{fig:cartoon}; more detailed pictures of the field structure of $\text{U}_\text{H}$ are provided in Ref.~\cite{Lopez23}.

\section{Apertured field special case}

Lastly, let us consider when the incident wavefield is passed through an aperture:
\begin{equation}
	\psi_\text{in}(0,y)
	= \Psi_\text{in}(0,y) \, \text{rect}\left( \frac{y}{W}\right)
\end{equation}

\noindent where $\Psi$ is the arbitrary un-apertured wavefield and $\text{rect}$ is the rectangular hat function of width $W$. If the aperture width is made infinitesimally small, then one can show that Eq.~\ref{eq:solUH} yields
\begin{equation}
	\lim_{W \to 0}
	\psi(x, y)
	= W N \, \Psi_\text{in}(0,0)
	\text{U}_\text{H}
	\left(
		\sqrt[3]{3} \frac{x - L}{\delta}
		,
		\frac{ 
			y
		}{\sqrt[6]{3} \, \delta}
		,
		- \frac{\lambda L}{2\pi \sqrt[3]{3} \, \delta^2 }
	\right)
	.
\end{equation}

\noindent Hence, an aperture drives all incident fields to an unfocused hyperbolic umbilic function. The journey of a given incident field to this common final state can be complicated, however. Specific examples for an incident plane and Gaussian-focused waves are given in Ref.~\cite{Lopez23}.

\section{Discussion \& Conclusion}

Here I have summarized a new description for general wavefields near turning points, illustrated with a number of simple examples. These results can be useful for developing improved reduced models for wave processes. For example, ECRH during the rampup phase on spherical tokamaks~\cite{Lopez18a} will invariably encounter the critical focusing case as the cutoff layer surpasses the resonance layer; applying caustic-agnostic methods such as metaplectic geometrical optics~\cite{Lopez20,Lopez21,Lopez22,Lopez22t} to such ray-tracing simulations will ensure that the hyperbolic umbilic function is calculated correctly.

\end{document}